\documentclass[11pt]{article}
\usepackage{amssymb, latexsym, amsmath}
\usepackage{amsfonts, graphics}
\def\R{\mathbb R}

\def\be{\begin{equation}}
\def\ee{\end{equation}}
\def\bea{\begin{eqnarray}}
\def\eea{\end{eqnarray}}
\def\beas{\begin{eqnarray*}}
\def\eeas{\end{eqnarray*}}

\def\open#1{\setbox0=\hbox{$#1$}
\baselineskip = 0pt
\vbox{\hbox{\hspace*{0.3 \wd0}\tiny $\circ$}\hbox{$#1$}}
\baselineskip = 11pt\!}

\newcommand{\prfe}{\hspace*{\fill} $\Box$

\smallskip \noindent}

\begin{document}
\sloppy
\newtheorem{theorem}{Theorem}[section]
\newtheorem{definition}[theorem]{Definition}
\newtheorem{proposition}[theorem]{Proposition}
\newtheorem{example}[theorem]{Example}
\newtheorem{remark}[theorem]{Remark}
\newtheorem{cor}[theorem]{Corollary}
\newtheorem{lemma}[theorem]{Lemma}

\renewcommand{\theequation}{\arabic{section}.\arabic{equation}}

\title{Spherically symmetric equilibria for
       self-gravitating kinetic or fluid models in the non-relativistic
       and relativistic case---A simple proof
       for finite extension}

\author{Tobias Ramming, Gerhard Rein\\
        Fakult\"at f\"ur Mathematik, Physik und Informatik\\
        Universit\"at Bayreuth\\
        D-95440 Bayreuth, Germany\\
        email: tobias.ramming@uni-bayreuth.de\\
        \phantom{ema}gerhard.rein@uni-bayreuth.de} 

\maketitle
\begin{abstract}
We consider a self-gravitating collisionless gas
as described by the Vlasov-Poisson or Einstein-Vlasov system
or a self-gravitating fluid ball as described by the
Euler-Poisson or Einstein-Euler system. We give a simple
proof for the finite extension of spherically symmetric
equilibria, which covers all these models simultaneously.
In the Vlasov case the equilibria are characterized by a 
local growth condition on the microscopic equation of state, 
i.e., on the dependence of the particle
distribution on the particle energy, at the cut-off energy $E_0$,
and in the Euler case by the corresponding growth condition
on the equation of state $p=P(\rho)$ at $\rho=0$.
These purely local conditions are slight generalizations to known
such conditions.
\end{abstract}
\section{Introduction}
\setcounter{equation}{0}
In astrophysics, matter distributions which interact by gravity
arise on many different scales. While the Euler-Poisson system 
can be used as a simple model for a single star, a large ensemble of
stars such as a galaxy or globular cluster where collisions
among the stars are sufficiently rare to be neglected
is typically modeled
by the Vlasov-Poisson system; both systems are non-relativistic
and possess relativistic counterparts. We refer to \cite{BT,C} for 
astrophysical background of these systems.
In a well known approach
to constructing corresponding equilibrium solutions,
the system under investigation is---by a suitable ansatz---reduced 
to a semi-linear
elliptic equation for the potential or its relativistic analogue.
The crucial question then is, under which assumptions on the ansatz
the resulting
steady state has finite mass and compact support, since only 
such states are of possible interest from a physics point of view.
In the present paper we give a simple proof
for these finiteness properties which works for all the indicated models
simultaneously and covers (and slightly extends) all those cases
known from the literature where the assumption is purely local at
the cut-off energy or at $\rho=0$ respectively.

In order to be more precise we first consider the models where
matter is described as a collisionless gas; for the necessary details
of what we outline below we refer to the next section. 
In the non-relativistic and time independent case the ensemble
of particles (stars) is described by its
density on phase space, $f=f(x,v)\geq 0,\ x, v \in \R^3$,
which obeys the Vlasov-Poisson system 
\be \label{vlasov}
v \cdot \nabla_x f - \nabla U \cdot \nabla _v f=0,
\ee
\be \label{poisson}
\Delta U = 4 \pi \rho,\ \lim_{|x|\to \infty} U(x)=0, 
\ee
\be \label{rhodef}
\rho (x) = \int f(x,v)\,dv. 
\ee
Here $U=U(x)$ denotes the gravitational potential
and $\rho$ the spatial mass density induced by $f$; we assume that 
all the particles have the same mass which we normalize to unity. 
Clearly, the particle energy
\be \label{parten}
E=E(x,v)=\frac{1}{2} |v|^2 + U(x)
\ee
satisfies the Vlasov equation (\ref{vlasov}), and hence any
function of the form
\be \label{isoansatz}
f=\phi(E)
\ee
with a suitable, prescribed function $\phi$ does as well.
The time independent Vlasov-Poisson system thus is reduced
to the semi-linear Poisson equation
\be \label{poissonsl}
\Delta U = 4 \pi \int\phi\left(\frac{1}{2} |v|^2 + U\right)\,dv,
\ \lim_{|x|\to \infty} U(x)=0, 
\ee
and the question is under what conditions on $\phi$ the latter
equation has solutions and whether the resulting steady states
have finite mass and compact support. A necessary condition
for the latter is that $\phi(E)=0$ for $E>E_0$ where $E_0$
is a suitable cut-off energy, cf.~\cite[Thm.~2.1]{RR00}.

If instead we describe the matter as an ideal, compressible fluid,
then all that remains of the Euler equations in the static, time independent
case is the equation
\be \label{euler}
\nabla p + \rho \nabla U = 0,
\ee
where the pressure $p$ depends on the mass density $\rho$
via an equation of state
\be \label{eqofst}
p=P(\rho),
\ee 
and the gravitational potential obeys the Poisson equation
(\ref{poisson}). If $P$ is strictly increasing on $[0,\infty[$
then (\ref{euler}) and (\ref{eqofst}) can be used to express
$\rho$ as a function of $U$, and again the system is reduced to
a semi-linear Poisson equation.

The results in \cite{GNN}
imply that solutions to these semi-linear Poisson problems
which lead to finite mass and compact support must be 
spherically symmetric. Hence we do not loose any relevant
equilibria if we make this assumption from the start.
Under this assumption the characteristic flow of the
Vlasov equation has the additional invariant
\be \label{Ldef}
L := |x\times v|^2,
\ee
the modulus of angular momentum squared. 
We generalize the ansatz (\ref{isoansatz}) to
\be \label{ansatz}
f= \phi(E) L^l 
\ee
which allows for a certain anisotropy in the Vlasov case; here $l>-1/2$.

Suppose now that we wish to describe the analogous physical systems
in a relativistic set-up.  On the kinetic level we can consider
the so-called relativistic Vlasov-Poisson system,
where the Vlasov equation is changed to 
\be \label{srvlasov}
\frac{v}{\sqrt{1+|v|^2}} \cdot \nabla_x f - \nabla U \cdot \nabla _v f=0,
\ee
while the Poisson equation (\ref{poisson}) and (\ref{rhodef})
remain unchanged; like all other physical constants the speed of light
is normalized to unity. The particle energy is redefined as
\be \label{srparten}
E=E(x,v)=\sqrt{1+|v|^2} + U(x),
\ee
and the same reduction procedure as outlined above applies.
For the static Euler equations the velocity field
is identically zero, and hence there is no difference between
the Euler-Poisson and the relativistic Euler-Poisson systems here.

The relativistic Vlasov-Poisson system is neither Galilei nor
Lorentz invariant and 
is only included here to show that our 
simple proof covers all the models of this type. The genuinely
relativistic case has to be modeled in the context of general
relativity. Assuming spherical symmetry we use Schwarzschild 
coordinates and write the metric in the form
\[
ds^2 = - e^{2\mu}dt^2 +e^{2\lambda} dr^2 + 
r^2 \left(d\theta^2 +\sin^2 \theta d \varphi ^2\right),
\]
where Schwarzschild time $t$ coincides with the proper time
of an observer who is at rest at spatial infinity, $r\geq0$
is the area radius, and the polar angles $\theta\in [0,\pi]$
and $\varphi\in [0,2\pi]$ coordinatize the orbits of symmetry.
The static Einstein-Vlasov system takes the form
\be \label{grvlasov}
\frac{v}{\sqrt{1+|v|^2}} \cdot \nabla_x f - \sqrt{1+|v|^2} \mu'
\frac{x}{r} \cdot \nabla_v f=0, 
\ee
\be \label{grfield1}
e^{-2\lambda} \left(2r \lambda'-1\right) +1 = 8 \pi r^2 \rho,
\ee
\be \label{grfield2}
e^{-2\lambda} \left( 2r \mu'+1\right) -1 = 8 \pi r^2 p, 
\ee
\be \label{grrhodef}
\rho(r) = \rho(x) = \int \sqrt{1+|v|^2}\ f(x,v)\, dv,
\ee
\be \label{grpdef}
p(r) = p(x) = \int \left( \frac{x\cdot v}{r} \right)^2 f(x,v)
\frac{dv}{\sqrt{1+|v|^2}}. 
\ee
Here $x = \left( r \sin \theta \cos \varphi , 
r \sin \theta \sin\varphi, r\cos \theta\right)$ 
so that $r=|x|$ is the
Euclidean norm of $x\in \R^3$, $\cdot$ denotes the
Euclidean scalar product on $\R^3$,
$f$ is spherically symmetric,
i.e., $f(x,v)=f(Ax,Av)$ for all $A\in \mathrm{SO}(3)$,
and $'$ denotes the derivative with respect to $r$.
As to the choice of the momentum variable $v\in\R^3$
which leads to
the above form of the system we refer to \cite{Rein95,RR92}. 
As boundary conditions we require asymptotic flatness, i.e., 
\be \label{asflat}
\lim_{r\rightarrow\infty} \mu (r) = 
\lim_{r\rightarrow \infty} \lambda (r) = 0,
\ee
and a regular center, i.e., 
\be \label{regcenter}
\lambda (0) = 0. 
\ee
For the static Einstein-Vlasov system the particle energy takes the
form
\be \label{grparten}
E=E(x,v)=e^{\mu(x)}\sqrt{1+|v|^2},
\ee
and an ansatz of the form (\ref{ansatz}) again satisfies the
corresponding Vlasov equation. With this ansatz the quantities $\rho$
and $p$ defined in (\ref{grrhodef}) and (\ref{grpdef})
become functions of $\mu$. Moreover, $e^{-2\lambda}$ can, via
(\ref{grfield1}) and (\ref{regcenter}), be expressed 
in terms of $\rho$,
and the whole system is reduced to a single equation for $\mu$
which arises from (\ref{grfield2}).

To complete the set of models which we consider we turn to the 
Einstein-Euler system. Since we consider the static case, 
the Euler equations reduce to the single equation
\be \label{greuler}
\nabla p + (p+\rho)\, \nabla \mu = 0,
\ee
where the pressure $p$ depends on $\rho$
via an equation of state like (\ref{eqofst}). The field
equations (\ref{grfield1}), (\ref{grfield2}) together with
the boundary conditions (\ref{asflat}), (\ref{regcenter})
remain unchanged. As in the case of the Euler-Poisson system
$\rho$ becomes a function of $\mu$, and the system is reduced
to a single equation for $\mu$.

Up to technical requirements we make the following assumptions. 
In the kinetic case,
\[
\phi(E) \geq c (E_0 - E)^k
\] 
for $E<E_0$ close to the cut-off energy $E_0$, where $c>0$
and $k<l+3/2$. For the fluid case we require
that $P(\rho)$ is strictly increasing for $\rho>0$ with
\[
P'(\rho)\leq c \rho^{1/n}
\]
for $\rho>0$ and small, where $0<n<3$. In passing we remark that
if one computes the pressure induced by the ansatz (\ref{ansatz})
in the isotropic case $l=0$ then it can be written
as a function of $\rho$ which satisfies the fluid case assumption
with $n=k+3/2$, and the restrictions on the growth rates fit.

We compare our result with known results from the literature.
The classical example in the context of the Vlasov-Poisson
system are the polytropic models where
\be \label{polytr}
f(x,v) = (E_0 -E)_+^k L^l;
\ee
the subscript $+$ denotes the positive part. 
The resulting semi-linear Poisson equation is the Lane-Emden-Fowler
equation. Based on the analysis in \cite{Sans} the corresponding 
steady states were analyzed in \cite{BFH}. Here $k, l > -1$ with
$k+l+3/2 \geq 0$. Compactly supported steady states are
obtained for $k < 3 l + 7/2$, for $k=3 l + 7/2$ the mass is still
finite but the support is $\R^3$, and for  $k>3 l + 7/2$ the mass
becomes infinite. In \cite{Rein94,RR93} extensions of these and
related results were given for the Einstein-Vlasov system.
In \cite{RR00} it was shown both for the Vlasov-Poisson
and the Einstein-Vlasov systems that a sufficient condition
for a compact support is that the ansatz is 
of the form (\ref{polytr}) asymptotically for $E \to E_0$, 
but this purely local
condition is sufficient only if $k<l+3/2$. This result was motivated
by and relied on a corresponding analysis for the Einstein-Euler system
\cite{M}. In \cite{RR00} a list of examples from the
astrophysics literature is given which are covered by such a
purely local condition, and it is also demonstrated by a
suitable counterexample that such purely local methods
fail for $k>l+3/2$. The analysis in the present paper relies on
a purely local characterization of the (microscopic) equation
of state as well and is subject to the same restriction.
Results which are not subject to this restriction  have been 
investigated by quite sophisticated dynamical
systems methods in \cite{FHU,HRU,HRoeU,HU}. Using a global characterization 
of the ansatz function $\phi$ these results cover
polytropes for the full range of exponents mentioned above
under a size restriction on the initial data.
Our analysis is much closer to the ones in \cite{M,RR00},
our conditions are less restrictive in that only an estimate
and not an asymptotic behavior is needed at the cut-off,
but more important from our point of view is that our proof
is transparent and short---cf.~Section~3---, and it applies
to all the different systems specified above.

There is by now a rich literature on the stability of
steady states of the Vlasov-Poisson system, cf.\
\cite{GR,HR,LMR,Rein07} and the references there.
It is interesting that the character of the stability analysis changes
at the threshold $k=3/2$---$k$ is again a growth rate
for the ansatz function and $l=0$ here---in the sense that below this
threshold one can use a reduction procedure which gives
a stability result simultaneously for the Vlasov-Poisson and
Euler-Poisson systems while such an approach does not work
above this threshold \cite{Rein02,Rein03,Rein07}.
We refer to \cite{Jang} for a complementary instability
result in the Euler-Poisson case.

The paper proceeds as follows. In the next section we show
in more detail how in the static case the models
we consider can be reduced to a single equation
for a suitably defined function $y$, related to either $U$ or $\mu$.
The arguments there are known, but 
we need to put them into a common
framework. The steady state under consideration has compact support
if and only if the function $y$, which starts with a positive value
at the origin and is decreasing, has a zero. It turns out that in all
the cases considered, $y$ satisfies an inequality of the form
\be \label{master}
y'(r) \leq -\frac{m(r)}{r^2},
\ee
where the mass function $m$ is defined in terms of $\rho$ by
\[
m(r)= 4\pi \int_0^r s^2 \rho(s)\,ds,
\]
and $\rho(r)=r^{2l} g(y(r))$ is given in terms of $y$. 
In Section~3 we prove that under a condition
on the behavior of $g$ at $y=0$, all functions which satisfy (\ref{master})
have a zero. This rests on two simple observations. Firstly, the mass function
is increasing since the mass-energy density $\rho$ is non-negative,
and secondly, the latter function is in the present context
always decreasing, up to the possible anisotropy factor $r^{2l}$
in the Vlasov case.
In the last section we translate the general condition from
Section~3 into a condition on the microscopic equation of state
$\phi$ or the macroscopic equation of state $P$ respectively
and obtain the compact support property for all the models considered
above.

\section{The basic set-up}
\setcounter{equation}{0}
In this section we discuss in more detail how the analysis of steady states
for the systems 
which were introduced above can be reduced
to that of a suitable master equation for the potential or a 
related quantity. 

\subsection{Kinetic models}

\subsubsection{The Vlasov-Poisson system}\label{vpsetup}

In the ansatz (\ref{ansatz}) a cut-off energy $E_0$ has to be specified
which is the value of the potential at the boundary
of the support of the matter. On the other hand we have the standard 
boundary condition in (\ref{poisson}) at infinity, and due to spherical 
symmetry is seems natural to parametrize for a fixed
ansatz function $\phi$  the solutions  
by prescribing the value $U(0)$ of the 
potential at the center. Since this is one free parameter respectively one
condition too many it is natural to slightly modify the ansatz
(\ref{ansatz}). We prescribe a function $\Phi$ and
make the ansatz
\be \label{trueansatz}
f(x,v)= \Phi(E_0 - E) L^l
\ee
with $l>-1/2$, and we look for $y=E_0-U$ with a prescribed value at the
origin, $y(0)=\open{y}>0$. Once a solution $y$ with a zero is 
found we define $E_0:= \lim_{r\to \infty} y(r)$ and $U:= E_0 -y$.
In this way the cut-off energy $E_0$ is eliminated as a free
parameter and becomes part of the solution. The following technical
assumption on $\Phi$ is required for the reduction procedure,
but it does in general not guarantee the compact support
of the resulting steady states.

\noindent
{\bf Assumptions on $\Phi$.} 
$\Phi:\R \to [0,\infty[$ is measurable,
$\Phi (\eta) = 0$ for $\eta < 0$, and 
$\Phi > 0$ a.\ e.ß on some interval $[0,\eta_1]$ with $\eta_1>0$.
Moreover, there exists $\kappa > -1$ such that
for every compact
set $K\subset \R$ there exists a constant $C > 0$ such that
\[
\Phi(\eta) \leq C \eta^\kappa,\ \eta \in K.
\]
If we substitute the ansatz (\ref{trueansatz}) into the definition 
(\ref{rhodef}) of $\rho$ we find after a short computation that
for $U(r) < E_0$,
\beas
\rho(r)
&=&
c_l r^{2l} \int_{U(r)}^{E_0} \Phi(E_0-E)\, \left(E-U(r)\right)^{l+1/2} dE\\
&=&
c_l r^{2l} \int_0^{E_0 - U(r)} \Phi(\eta)\, 
\left(E_0 - U(r) -\eta\right)^{l+1/2} d\eta
\eeas
and $\rho (r)=0$ if $U(r) \geq E_0$. Here 
\[
c_l:= 2^{l+3/2} \pi \int_0^1 \frac{s^l}{\sqrt{1-s}} ds.
\]
Hence in terms of $y:= E_0-U$ we find that
\be \label{rhoyrel}
\rho(r) = r^{2 l} g(y(r))
\ee
where
\be \label{vpgdef}
g(y):= \left\{\begin{array}{ccl}
c_l \int_0^y \Phi(\eta)\, (y -\eta)^{l+1/2} d\eta&,&y>0,\\
0 &,& y\leq 0.
\end{array}\right.
\ee
Under the above assumptions on $\Phi$ it follows by
Lebesgue's dominated convergence theorem that
$g\in C(\R)\cap C^1(]0,\infty[)$ with
\[
g'(y) = (l+1/2)\, c_l 
\int_0^y \Phi(\eta)\, (y -\eta)^{l-1/2} d\eta,\ y>0,
\]
and $g\in C^1(\R)$ if $\kappa + l +1/2 >0$.
Due to spherical symmetry
the semi-linear Poisson equation (\ref{poissonsl})
can in terms of $y$ be written as 
\be \label{yeq2ndo}
\frac{1}{r^2} \left(r^2 y'\right)' = 
- 4 \pi r^{2l} g(y).
\ee
In terms of the Cartesian variables we want
potentials $U\in C^2(\R^3)$ i.e., $y\in C^2(\R^3)$. Hence
we require that $y'(0)=0$,
integrate (\ref{yeq2ndo}) once and have to solve the equation
\be\label{yeq}
y'(r) = - \frac{m(r)}{r^2}
\ee
where 
\be \label{mdef}
m(r) = m(r,y) = 4\pi \int_0^r s^{2l+2} g(y(s))\, ds.
\ee
For any $\open{y} >0$ the equation (\ref{yeq}) has a unique solution
$y\in C^1([0,\infty[)$ with $y(0)=\open{y}\,$, and we briefly review
the proof.
Firstly, a standard contraction argument shows that there is a unique,
local solution on some short interval $[0,\delta[$. This solution extends
uniquely to a maximal interval $[0,r_{\mathrm{max}}[$ where by monotonicity,
$0<y(r) < \open{y}\,$. If  $r_{\mathrm{max}}=\infty$, we are done,
if not, then necessarily $y(r_{\mathrm{max}})=0$, and again by monotonicity,
$y$ uniquely extends to the right via
\[
y'(r) = - \frac{m(r_{\mathrm{max}})}{r^2},\ r > r_{\mathrm{max}}.
\]
In addition, $y'(0)=0$, and the regularity of $g$ implies that
$y\in C^2(\R^3)$ as desired. 

In the next section we specify a condition on $g$ which guarantees that
the solution $y$ has a zero at some finite radius $R$. In the last section 
we translate that condition into one on the
ansatz function $\Phi$, and from $y$ and $\Phi$
we then generate a steady state
of the Vlasov-Poisson system which in space is supported on the ball
with radius $R$ centered at the origin.

\subsubsection{The relativistic Vlasov-Poisson system} \label{rvpsetup}

We make the same ansatz (\ref{trueansatz}) as for the Vlasov-Poisson system,
with a function $\Phi$ which has the same properties as in \ref{vpsetup},
but the particle energy is now defined by (\ref{srparten}).
We again reduce the full system to the equation (\ref{yeq}),
but now for $y=E_0 -U-1$; note that by (\ref{srparten}),
$E \geq 1+ U(r)$. 
The mass function $m$ is defined as in (\ref{mdef}), but in the relation
(\ref{rhoUrel}) we obtain a different form for the function $g$:
\be \label{rvpgdef}
g(y):= \left\{\begin{array}{ccl}
c_l \int_0^y \Phi(\eta)\, (1+y -\eta)\, 
\left((1+y -\eta)^2-1\right)^{l+1/2} d\eta&,&y>0,\\
0 &,& y\leq 0,
\end{array}\right.
\ee
where
\[
c_l := 2 \pi \int_0^1 \frac{s^l}{\sqrt{1-s}} ds.
\]
The function $g$ looks more complicated now, but it has the
same properties which were stated in the Vlasov-Poisson case,
and we arrive at the same type of set-up as in \ref{vpsetup}.

\subsubsection{The Einstein-Vlasov system}

First we observe that the unique solution to the field equation
(\ref{grfield1}) which satisfies the boundary condition (\ref{regcenter})
is given by
\be \label{lambdamrel}
e^{-2\lambda(r)} = 1-\frac{2 m(r)}{r},
\ee
where the mass function $m$ is defined as above. This relation
defines $\lambda$ only as long as the right hand side is positive,
a restriction which is due to the fact that Schwarzschild coordinates cannot
cover regions of spacetime which contain a trapped surface.
If we eliminate $\lambda$ via (\ref{lambdamrel}) and observe that the
particle energy (\ref{grparten}) depends only on $\mu$, 
the static Einstein-Vlasov system is reduced
to a single equation for $\mu$, namely to (\ref{grfield2}).

In order to arrive at a master equation for a suitable quantity $y$
which is qualitatively of the same form as before we need to
adapt the ansatz to the fact that the particle energy (\ref{grparten}) 
is no longer the sum of
a kinetic and a potential part. Hence we make the ansatz that
\be \label{grtrueansatz}
f(x,v) = \Phi\left(1-\frac{E}{E_0}\right)\, L^l,
\ee
where $\Phi$ has the properties stated in \ref{vpsetup}.
We define $y:= \ln E_0 - \mu$ so that $e^\mu = E_0/e^y$; notice that
the particle energy (\ref{grparten}) is always positive 
so we require that $E_0>0$. If we substitute the above ansatz
into the definitions (\ref{grrhodef}) and (\ref{grpdef})
we obtain the relations
\be \label{grrhoyrel}
\rho(r) = r^{2l} g(y(r)), \quad
p(r) = r^{2l} h(y(r)), 
\ee
where
\be \label{evgdef}
g(y) := \left\{\begin{array}{ccl}
c_l e^{3 y} \int_0^{1-e^{-y}} \Phi(\eta)\, (1-\eta)^2\, 
\left(e^{2y}(1-\eta)^2-1\right)^{l+1/2} d\eta&,&y>0,\\
0 &,& y\leq 0,
\end{array}\right.
\ee
with $c_l$ as in \ref{rvpsetup}, and
\be \label{evhdef}
h(y) := \left\{\begin{array}{ccl}
d_l e^{y} \int_0^{1-e^{-y}} \Phi(\eta)\,  
\left(e^{2y}(1-\eta)^2-1\right)^{l+3/2} d\eta&,&y>0,\\
0 &,& y\leq 0,
\end{array}\right.
\ee
with
\[
d_l := 2 \pi \int_0^1 s^l \sqrt{1-s}\, ds.
\]
The functions $g$ and $h$ have the same regularity properties
as the function $g$ in \ref{vpsetup}, cf.\ \cite[Lemma 2.2]{RR00}, 
and the static Einstein-Vlasov
system is reduced to the equation
\be \label{gryeq}
y'(r)= - \frac{1}{1-2 m(r)/r} \left(\frac{m(r)}{r^2} + 4 \pi r p(r)\right) ,
\ee
where the mass function $m$ is defined in terms of $\rho$ as
before and $\rho$ and $p$ are given in terms of $y$ by (\ref{grrhoyrel}). 
For any $\open{y} >0$
there exists a unique solution $y\in C^1([0,\infty[)$ of
(\ref{gryeq}) with $y(0)=\open{y}\,$
which due to the issue of the positivity of the denominator
is less easy to see, cf.\ \cite{Rein94,RR93}.

Once a solution to (\ref{gryeq}) is obtained, we define 
$y_\infty := \lim_{r\to \infty} y(r)$, $E_0:=e^{y_\infty}$,
and $\mu = \ln E_0 -y$. Together with the ansatz (\ref{grtrueansatz})
this yields a steady state with all the desired properties,
provided $y$ has a zero. It turns out that in order to show
the latter not the full information of (\ref{gryeq})
is needed, but only the following inequality which immediately 
follows from that equation:
\be \label{yineq}
y'(r) \leq - \frac{m(r)}{r^2}.
\ee
The difference between
the Newtonian, special relativistic, or general relativistic cases
is then reflected only in the definition of the function $g$,
and only an estimate for $g$ at $y=0$ which holds in all three cases
is needed to guarantee a zero for $y(r)$.

\subsection{Fluid models}

\subsubsection{The Euler-Poisson system} \label{epsetup}

We use the static Euler equation (\ref{euler})
together with the equation of state (\ref{eqofst}) in order
to express $\rho$ in terms of $U$.

\noindent
{\bf Assumptions on $P$.}
Let $P\in C^1([0,\infty[)$ be such that $P'>0$ on $]0,\infty[$, and
\[
\int_0^1 \frac{P'(s)}{s}ds < \infty .
\]

\noindent
We define
\[
Q(\rho):= \int_0^\rho \frac{P'(s)}{s}ds,\ \rho\geq 0,
\]
so that $Q\in C([0,\infty[)\cap C^1(]0,\infty[)$ with $Q(0)=0$
and $Q'(\rho)= P'(\rho)/\rho$ for $\rho>0$.
When written in the radial variable $r$, (\ref{euler}) reads
\be \label{eulerrad}
P'(\rho)\, \rho' + \rho U' = 0.
\ee
If we divide by $\rho$, integrate with respect to $r$ and apply
a change of variables it turns out that the
pair $(\rho,U)$ satisfies 
(\ref{eulerrad})---at least where $\rho(r)>0$---, provided
\be \label{rhoUrel}
Q(\rho(r)) = c - U(r),\ r\geq 0,
\ee
with some integration constant $c$ which like the cut-off energy $E_0$ 
above is the value of the potential at the boundary of the matter support.
Let
\[
y_{\mathrm{max}} := \int_0^\infty \frac{P'(s)}{s}ds 
= \lim_{\rho\to \infty} Q(\rho) \in ]0,\infty].
\]
Then $Q:[0,\infty[ \to [0,y_{\mathrm{max}}[$ is one-to-one and onto,
and we define
\be \label{epgdef}
g(y):=\left\{
\begin{array}{ccl}
Q^{-1}(y) &,& 0<y<y_{\mathrm{max}},\\
0&,&y\leq 0.
\end{array} \right.
\ee
Then $g\in C(]-\infty,y_{\mathrm{max}}[)\cap C^1(]0,y_{\mathrm{max}}[)$,
and writing $y=c-U$ we invert the relation (\ref{rhoUrel})
to read as in (\ref{rhoyrel}) with $l=0$ there.
Hence the static Euler-Poisson system is reduced to the same equation
(\ref{yeq}) with mass function defined by (\ref{mdef}) with $l=0$
and with $g$ defined by (\ref{epgdef}) instead of by (\ref{vpgdef}).
Hence we are in exactly the same situation as in the Vlasov-Poisson
case in that the crucial question is whether $y$ has a zero
$R$.

\subsubsection{The Einstein-Euler system} \label{eesetup}

Similarly to \ref{epsetup}
we use the static Euler equation (\ref{greuler})
together with the equation of state (\ref{eqofst}) in order
to express $\rho$ in terms of $\mu$.

\noindent
{\bf Assumptions on $P$.}
Let $P\in C^1([0,\infty[),\ P\geq 0$ be such that $P'>0$ on $]0,\infty[$, and
\[
\int_0^1 \frac{P'(s)}{s+P(s)}ds < \infty .
\]
We define
\[
Q(\rho):= \int_0^\rho \frac{P'(s)}{s+P(s)}ds,\ \rho\geq 0,
\]
so that $Q\in C([0,\infty[)\cap C^1(]0,\infty[)$ with $Q(0)=0$
and $Q'(\rho)= P'(\rho)/(\rho+P(\rho))$ for $\rho>0$.
We rewrite (\ref{greuler}) in the radial variable $r$,
\be \label{greulerrad}
P'(\rho)\, \rho' + (\rho+ P(\rho))\, \mu' = 0.
\ee
If we divide by $\rho+P(\rho)$, integrate with respect to $r$ and apply
a change of variables we see that the
pair $(\rho,\mu)$ satisfies 
(\ref{greulerrad})---at least where $\rho(r)>0$---, provided
\be \label{rhomurel}
Q(\rho(r)) = c - \mu(r),\ r\geq 0,
\ee
with some integration constant $c$.
Let
\[
y_{\mathrm{max}} := \int_0^\infty \frac{P'(s)}{s+P(s)}ds 
= \lim_{\rho\to \infty} Q(\rho) \in ]0,\infty].
\]
Then as before $Q:[0,\infty[ \to [0,y_{\mathrm{max}}[$ is one-to-one and onto,
and we define $g$ by (\ref{epgdef})
which has the same regularity properties as before. 
Writing $y=c-\mu$ we can invert the relation (\ref{rhomurel})
to read as in (\ref{rhoyrel}) with $l=0$ there.
Hence the static Einstein-Euler system is reduced to the same equation
(\ref{gryeq}) with mass function defined by (\ref{mdef}) with $l=0$.
The issue again is whether $y$ has a zero,
and this will be determined by the inequality (\ref{yineq}).

\section{The compact-support-Lemma}
\setcounter{equation}{0}
The key to the compact support property for all the models discussed
above is the following result.

\begin{lemma} \label{keylemma}
Let $y\in C^1([0,\infty[)$ with $y(0)=\open{y} \in ]0,y_{\mathrm{max}}[$ 
satisfy the estimate
\[
y'(r) \leq -\frac{m(r)}{r^2} \ \mbox{on}\ [0,\infty[,
\]
where
\[
m(r)= m(r,y) := 4\pi \int_0^rs^{2+2l} g(y(s))\, ds,
\]
$g\in C(]-\infty,y_{\mathrm{max}}[)$ is increasing with
$g(y)=0$ for $y \leq 0$ and $g(y)>0$ for $y >0$,
and $l>-1/2$. Let $g$ satisfy the estimate
\[
g(y) \geq c \,y^{n+l}\ \mbox{for}\ 0<y<y^\ast
\]
with parameters $c>0$, $y^\ast >0$, and $0<n<3+l$.
Then the function $y$ has a unique zero.
\end{lemma}

\noindent{\bf Proof.} 
Since $y$ is decreasing, the limit 
$y_\infty := \lim_{r\to\infty} y(r)\in [-\infty,\infty[$ exists, and 
we need to show that $y_\infty < 0$. This will imply the existence
of a zero of $y$ which will be unique by the strict monotonicity 
of this function. Assume that
$y_\infty >0$. Then $y(r) \geq y_\infty$ on $[0,\infty[$, and by 
the monotonicity of $g$,
\[
m(r) \geq  4\pi g(y_\infty) \int_0^rs^{2+2l}  ds 
= \frac{4\pi}{2l+3} g(y_\infty)\,r^{2l+3}.
\]
If we put this into the estimate for $y'$ and integrate we obtain the
contradiction
\[
y(r) \leq \open{y} - C r^{2+2l} \to -\infty\ \mbox{as}\ r\to\infty;
\]
$C$ denotes a positive constant which may
depend on all the parameters, may change from line to line, but never 
depends on $r$. 

The argument so far is standard and well known, and the crucial task
is to derive a contradiction from the remaining possibility that 
$y_\infty=0$.
Firstly, we observe that $m$ is increasing in $r$ and positive
for $r>0$. Hence
\be \label{mest1}
m(r) \geq m(1)=:m_1 > 0\ \mbox{for}\ r\geq 1,
\ee
and
\be \label{yest1}
y(r) = - \int_r^\infty y'(s)\, ds \geq  m_1 \int_r^\infty \frac{ds}{s^2}
= \frac{m_1}{r}\ \mbox{for}\ r\geq 1.
\ee
Secondly, since $g$ is increasing and $y$ decreasing,
\be \label{mest2}
m(r) \geq  4\pi g(y(r)) \int_0^rs^{2+2l}  ds 
= \frac{4\pi}{2l+3} r^{2l+3} g(y(r)).
\ee
Hence 
\[
y'(r) \leq - \frac{4\pi}{2l+3} r^{2l+1} g(y(r)),\ r>0.
\]
By a simple change of variables this implies that for all $r>0$,
\[
\int^{\open{y}}_{y(r)} \frac{d\eta}{g(\eta)} = 
- \int_0^r \frac{y'(s)}{g(y(s))}ds 
\geq  \frac{4\pi}{2l+3} \int_0^r s^{2l +1} ds 
= \frac{4\pi}{(2l+3)(2l+2)} r^{2l+2}.
\]
Now we take $r>0$ sufficiently large so that $0<y(r)< y^\ast$;
recall that by assumption, $y(r)\to 0$ as $r\to\infty$.
Then by the growth assumption on $g$,
\[
C_1 r^{2l+2} 
\leq \int^{y^\ast}_{y(r)}\frac{d\eta}{g(\eta)} + C_2
\leq \frac{1}{c}\int^{y^\ast}_{y(r)}\frac{d\eta}{\eta^{n+l}} +C_2.
\]
We estimate the left hand side from below using (\ref{yest1}), multiply the
resulting estimate by $y(r)^{2l+2}$ and compute the integral where we
need to distinguish the cases $n+l\neq 1$ and $n+l=1$.
In the former case we find that
\[
C_1 \leq \frac{1}{c\,(1-l-n)} 
\left((y^\ast)^{1-l-n}- y(r)^{1-l-n}\right)\,y(r)^{2l+2} 
+ C_2 y(r)^{2l+2},
\]
in the latter
\[
C_1 \leq \frac{1}{c} \ln\left(\frac{y^\ast}{y(r)}\right)\,y(r)^{2l+2} 
+ C_2 y(r)^{2l+2},
\]
which holds for $r$ sufficiently large with positive constants $C_1,C_2$.
Since $2l+2 >0$ and $l+3-n>0$ and since by assumption $y_\infty=0$, 
the right hand side goes to zero 
as $r$ goes to infinity which is the desired contradiction. \prfe

\noindent
{\bf Remark.} The two
estimates (\ref{mest1}) and (\ref{mest2}) on which the above proof 
rests are quite obvious from a physics point of view.
The mass function
$m(r)$ is increasing in $r$ since the mass-energy density $\rho$ is
non-negative, and this yields (\ref{mest1}), and
$\rho$ is, up to a possible
anisotropy factor $r^{2l}$ in the Vlasov case, a decreasing function,
which yields (\ref{mest2}).
\section{Application to the various models}
\setcounter{equation}{0}
In this section we apply Lemma~\ref{keylemma} to the various models.
We have to check what type of ansatz function $\Phi$
or equation of state function $P$ leads to a relation
between the mass-energy density $\rho$ and the function $y$
with a functional dependence $g$ which satisfies the growth
condition in that lemma.

\subsection{Kinetic models}\label{kinappl}

Let $\Phi$ satisfy the assumptions stated in \ref{vpsetup}.
In addition let 
\be \label{phicond}
\Phi(\eta) \geq c \eta^k \ \mbox{for}\ \eta \in ]0,\eta_0[
\ee
for some parameters $c>0$, $\eta_0>0$, and $-1<k<l+3/2$.

\subsubsection{The Vlasov-Poisson system}\label{vpappl}

For $0<y<\eta_0$ the function $g$ defined by (\ref{vpgdef})
satisfies the following estimate; $C>0$ denotes a constant which can 
change from line to line and depends only on the parameters above:
\beas
g(y)
&\geq&
C \int_0^y  \eta^k (y-\eta)^{l+1/2}\, d\eta 
=
C y^{l+1/2}\int_0^y \eta^k (1-\eta/y)^{l+1/2}\, d\eta\\
&=&
C y^{k+l+3/2}\int_0^1 s^k (1-s)^{l+1/2}\, ds.
\eeas
Hence $g$ satisfies the assumption in Lemma~\ref{keylemma}
with $0<n=k+3/2 < 3+l$ by the assumption on $k$. 

\subsubsection{The relativistic Vlasov-Poisson system}\label{rvpappl}

In this case the corresponding function $g$ is defined in (\ref{rvpgdef}).
We observe that 
$(1+y-\eta)^2-1=(1+y-\eta+1)\,(1+y-\eta-1) \geq y-\eta$ and find that
\beas
g(y)
&\geq&
C \int_0^y  \eta^k (1+y-\eta)\,((1+y-\eta)^2-1)^{l+1/2}\, d\eta \\
&\geq&
C \int_0^y  \eta^k (y-\eta)^{l+1/2}\, d\eta \\
&=&
C y^{k+l+3/2}\int_0^1 s^k (1-s)^{l+1/2}\, ds.
\eeas
Again, $g$ satisfies the assumption in Lemma~\ref{keylemma}.

\subsubsection{The Einstein-Vlasov system}\label{evappl}

In this case the corresponding function $g$ is defined
in (\ref{evgdef}). We estimate analogously to \ref{rvpappl},
and in addition we observe that for $y$ sufficiently small,
$1-e^{-y} \geq y/2$ and $e^y \geq 1/2$. Hence
\beas
g(y)
&\geq&
C \int_0^{1-e^{-y}}  \eta^k \left((1-\eta)^2-e^{-2y}\right)^{l+1/2}\, d\eta \\
&\geq&
C \int_0^{1-e^{-y}}  \eta^k (1-\eta -e^{-y})^{l+1/2}\, d\eta \\
&=&
C (1-e^{-y})^{k+l+3/2}\int_0^1 s^k (1-s)^{l+1/2}\, ds
\geq
C y^{k+l+3/2}
\eeas
as desired.
We collect the results for the kinetic models into a theorem.
\begin{theorem} \label{kintheorem}
Let $\Phi$ satisfy the assumptions stated in \ref{vpsetup}
and (\ref{phicond}). Then for any $\open{y} > 0$ the reduced
equation (\ref{yeq}) or (\ref{gryeq}) has a unique solution
$y\in C^1([0,\infty[)$ with $y(0)=\open{y}$ which has a
unique zero $R>0$. By
\[
f(x,v) = \Phi\left(y(r) - \frac{1}{2} |v|^2\right) |x\times v|^{2l}
\]
or
\[
f(x,v) = \Phi\left(1+ y(r) - \sqrt{1+|v|^2}\right) |x\times v|^{2l}
\]
or
\[
f(x,v) = \Phi\left(1-e^{-y(r)}\sqrt{1+|v|^2} \right) |x\times v|^{2l}
\]
a static, spherically symmetric
solution to the Vlasov-Poisson or relativistic Vlasov-Poisson
or Einstein-Vlasov system is defined. This solution is compactly
supported, and its spatial support is the ball with radius $R$
centered at the origin. The parameter $\open{y}$ is related
to the potential $U$ or the metric quantity $\mu$ via
$\open{y}=U(R)-U(0)$ or $\open{y}=\mu(R)-\mu(0)$ respectively.
Moreover, $\rho, p \in C(\R^3)\cap C^1(B_{R}(0))$.
\end{theorem}

\subsection{Fluid models}\label{eulerappl}

Let $P$ satisfy the assumptions stated in \ref{epsetup}
or \ref{eesetup}.
In addition let 
\be \label{Pcond}
P'(\rho) \leq c \rho^{1/n} \ \mbox{for}\ \rho \in ]0,\rho_0[
\ee
for some parameters $c>0$, $\rho_0>0$, and $0<n<3$.
It turns out that in checking the condition for the corresponding
function $g$ we need not distinguish between the non-relativistic
and relativistic cases, since in both cases 
\[
Q(\rho) \leq \int_0^\rho \frac{P'(s)}{s} ds \leq C \rho^{1/n}
\]
for $0<\rho<\rho_0$. Since for positive arguments, $g$ is defined
as the inverse function to $Q$ this immediately yields the
estimate for $g$ required in Lemma~\ref{keylemma}, and we can 
sum up our results for the fluid case.
\begin{theorem} \label{fluidtheorem}
Let $P$ satisfy the assumptions stated in \ref{epsetup}
or \ref{eesetup}
and (\ref{Pcond}). Then for any 
$\open{y} \in ]0,y_{\mathrm{max}}[$ the reduced
equation (\ref{yeq}) or (\ref{gryeq}) has a unique solution
$y\in C^1([0,\infty[)$ with $y(0)=\open{y}$ which has a
unique zero $R>0$. By
\[
\rho = g(y),\ p=P(\rho)
\]
with $g$ defined in \ref{epsetup} or \ref{eesetup} respectively,
a static, spherically symmetric
solution to the Euler-Poisson or Einstein-Euler system is defined. 
This solution is supported  in the ball of radius $R$
centered at the origin. 
The parameter $\open{y}$ is related
to the potential $U$ or the metric quantity $\mu$ via
$\open{y}=U(R)-U(0)$ or $\open{y}=\mu(R)-\mu(0)$ respectively.
Moreover, $\rho \in C(\R^3)\cap C^1(B_{R}(0))$.
\end{theorem} 

\subsection{Final remarks}
\begin{enumerate}
\item
In the kinetic case it is straight forward to extend the above analysis to
an ansatz of the type
\[
f(x,v) =\Phi(E_0 -E)\, (L-L_0)_+^l
\]
or its general relativistic analogue,
where $L_0>0$ and the other parameters are as before.
Such an ansatz leads to steady states which have a vacuum region
at the center. This situation was investigated in
\cite{Rein99} by a perturbation argument, and the structure
of the resulting steady states was studied in \cite{AR}
by numerical means.

\item
The arguments from Lemma~\ref{keylemma} can easily be applied
to show that a solution to the equation (\ref{yeq2ndo})
with data $y(\open{r}\,)=\open{y}>0$, $y'(\open{r}\,)=\open{y}\,'$
prescribed at some radius $\open{r}>0$, has a zero to the right of
$\open{r}\,$, provided $\open{y}\,'<0$. The important point
is that again $y'(r) \leq - m(r)/r^2$ for $r\geq \open{r}$,
where $m(r) = 4\pi\int_{\open{r}}^r s^{2+2l} \rho(s)\, ds$.
This extension will be useful in \cite{Ramm2}.

\item
In the Euler case there is a size restriction on $\open{y}\,$,
if $y_{\mathrm{max}}<\infty$, i.e., if $P$ grows only sub-linearly
for large values of $\rho$. If the pressure is weak in this sense
it cannot support an arbitrarily large potential difference
between the center and the surface of the equilibrium matter
distribution. For equations of state which typically arise in physics
$P$ grows superlinearly for large values of $\rho$ so that
$y_{\mathrm{max}}=\infty$.
To see why no such restriction appears
in the kinetic case we consider for simplicity
the Vlasov-Poisson case with $l=0$. Then $p=h(y)=h(g^{-1}(\rho))$,
i.e., $P=h\circ g^{-1}$. A simple change of variables
shows that in this case
\[
Q(\rho)=
\int_0^\rho \frac{(h\circ g^{-1})'(s)}{s} ds =
\int_0^{g^{-1}(\rho)} \frac{h'(t)}{g(t)} dt \to \infty\ \mbox{as}\ 
\rho\to\infty 
\]
because in the isotropic Vlasov case $h'$ is a positive
multiple of $g$, cf.\ \cite[Lemma 2.2]{RR00}. Hence in the
Vlasov case $y_{\mathrm{max}}=\infty$.
\item
In the kinetic case, $\rho\in C^1(\R^3)$, provided
$\kappa + l + 1/2 >0$, and the same is true in the fluid
case under a suitable assumption on $P$.
\item
In the kinetic case the restriction $l>-1/2$ can be relaxed by
assuming more on $\Phi$.
For example, Lemma~\ref{keylemma} applies to all the polytropes 
(\ref{polytr}) in the Vlasov-Poisson case
with $k,l>-1$, $k+l+3/2>0$ and $k<l+3/2$,
since in that case $\rho(r)=c r^{2l} y(r)_+^{k+l+3/2}$.
\end{enumerate}
 
\noindent
{\bf Acknowledgment.}
The results reported here originate in the first author's doctoral
thesis \cite{Ramm}.

\end{document}